\documentclass[%
reprint,
superscriptaddress,
showpacs,preprintnumbers,
verbatim,
amsmath,amssymb,
aps,
pre,
]{revtex4-2}

\usepackage{color,mathtools}
\usepackage{graphicx}
\usepackage{dcolumn}
\usepackage{bm}
\usepackage{arydshln}
\usepackage{textcomp} 

\begin {document}

\title{Exact Generalized Langevin Dynamics of Pair Coordinates in Elastic Networks}


\author{Shunsuke Ando}%
\affiliation{%
  Department of Systems Engineering, Wakayama University, 
  Wakayama, 640-8510, Japan}

\author{Tomoya Urashita}%
\affiliation{%
  Department of Systems Engineering, Wakayama University, 
  Wakayama, 640-8510, Japan}

\author{Soya Shinkai}%
\affiliation{Laboratory for Developmental Dynamics, RIKEN Center for Biosystems
  Dynamics Research, Kobe 650-0047, Japan }%

\author{Tomoshige Miyaguchi}
\email{mygch@wakayama-u.ac.jp}
\affiliation{%
  Department of Systems Engineering, Wakayama University, 
  Wakayama, 640-8510, Japan} 


%
%


\date{\today}


\begin{abstract}

  Generalized Langevin equations (GLEs) provide a powerful framework for
  describing slow dynamics in soft-matter systems, but deriving an exact GLE for
  a reaction coordinate from an underlying many-body system remains generally
  difficult.  Here, we analytically derive an exact coupled GLE for the
  overdamped dynamics of two tagged beads in arbitrary harmonic elastic
  networks.  We further derive an exact GLE for their relative coordinate. The
  memory kernel and effective restoring force are expressed explicitly in terms
  of the network matrices, thereby providing a systematic reduction of the
  high-dimensional network dynamics to a pair coordinate. Within the
  small-displacement approximation, we further derive a GLE for the inter-bead
  distance, a central observable in distance-sensitive single-molecule
  experiments. These results therefore have broad potential applications in
  modeling proteins and other soft-matter systems.
  
\end{abstract}

\maketitle


\textit{Introduction.} Slow dynamics in complex many-body systems are often
described in terms of a small number of collective or reaction coordinates.
Such reduced descriptions are particularly important when the full dynamics span
a wide range of time scales, as in biomolecules \cite{yang03,ye18,li22} and
glass-forming liquids \cite{kob95,yamamoto98b}. In experiments and molecular
simulations, one often monitors only a few observables, such as tagged-particle
positions, end-to-end vectors, or intramolecular distances \cite{kuzmenkina05,
  kou04, min05, hu16,yamamoto21, ayaz21}. A central theoretical problem is
therefore to derive effective low-dimensional dynamics for such observables
directly from the underlying many-body system.


Memory effects are expected to play an essential role in such reduced
descriptions.  A natural framework is the generalized Langevin equation (GLE)
with a coordinate-independent memory kernel \cite{shimizu25}, which
characterizes the temporal nonlocality of the effective dynamics. Such GLEs have
been used to analyze slow relaxation in a variety of systems, ranging from
proteins \cite{kou04,ayaz21,dalton24} to gels \cite{azevedo25} and disordered
solids \cite{cui17}. However, deriving a GLE with a coordinate-independent
memory kernel for a chosen reaction coordinate from microscopic many-body
dynamics is generally nontrivial, because projection onto a reduced set of
coordinates can, in general, yield a GLE with a coordinate-dependent memory
kernel \cite{vroylandt22}. Throughout this Letter, unless otherwise stated, we use
the term GLE to refer to a GLE with a coordinate-independent memory kernel.


Among the observables of current interest, distance-like quantities are
particularly important because they are directly relevant to distance-sensitive
single-molecule experiments, including photoinduced electron transfer
\cite{kou04,min05} and F\"orster resonance energy transfer
\cite{kuzmenkina05}. For example, Min \textit{et al.} experimentally measured
distance fluctuations between a fluorescein--tyrosine pair within a protein
complex and showed that these fluctuations are well described by a GLE
\cite{min05}. Likewise, Ayaz \textit{et al.} analyzed all-atom
molecular-dynamics simulations of $\text{Ala}_9$ in water and found that an
averaged hydrogen-bond distance is well described by a GLE \cite{ayaz21}. On the
theoretical side, Xing \textit{et al.} studied the experiment of
Ref.~\cite{min05} using an elastic network model (ENM) constructed from a
protein structure in the Protein Data Bank \cite{xing06}.  Assuming that the
measured distance follows a GLE, they showed that the resulting memory kernel is
consistent with the experimental data, although this required friction
coefficients much larger than those estimated from the viscosity of water.
Moreover, Sain \textit{et al.} studied bond breaking in a linear polymer chain
by focusing on the two beads adjacent to the breaking bond \cite{sain06}.


These studies highlight the physical importance of distance observables, but it
remains unclear under what conditions an inter-bead distance or another reaction
coordinate obeys a GLE with a coordinate-independent memory kernel.  Exact
analytical results for coordinate-independent memory kernels are known only in
limited special cases, most notably for the end-to-end distance vector of the
phantom Rouse chain \cite{tian22}. This contrasts with single-bead motion, for
which GLEs have been derived for ideal network polymers
\cite{taloni10,miyaguchi24,durang24}.


\begin{figure}[b]
  
  \centerline{\includegraphics[width=8.5cm]{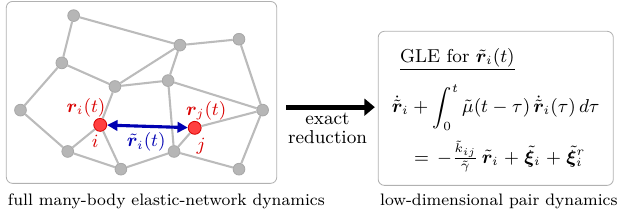}}
  
  \caption{\label{f.enm}Schematic illustration of the exact reduction from the
    full dynamics of an elastic network to a GLE for the relative coordinate
    $\tilde{\bm{r}}_i(t)=\bm{r}_i(t)-\bm{r}_j(t)$ of two tagged beads.  }
\end{figure}

In this Letter, we address this problem for dynamical ENMs. We derive an exact
GLE for the relative coordinate of two tagged beads in arbitrary dynamical ENMs
(Fig.~\ref{f.enm}) and, within the small-displacement approximation, a GLE for
the inter-bead distance. The corresponding memory kernel and effective restoring
force are obtained explicitly in terms of the network matrices, yielding a
systematic reduction from high-dimensional network dynamics to pair coordinates.


Elastic-network descriptions provide a natural setting for this problem because
they retain the network connectivity of the underlying many-body system while
remaining analytically tractable. Elastic network models were introduced as
coarse-grained models of proteins \cite{tirion96,bahar98} and have been shown to
describe the static fluctuations of folded globular proteins well
\cite{bahar98}. They have also been used to study dynamical properties of
proteins \cite{xing06,ciliberti06,caballero07,reuveni10,copperman15,togashi18,
  lapolla21,cecconi23,lam24,costantini25} and other network-forming soft-matter
systems such as chromatin \cite{shinkai20,yuan24} and cross-linked polymers and
gels \cite{gurtovenko98, gurtovenko05}. Here, we focus on the Gaussian network
model, whose harmonic structure allows an exact analytical treatment.



Our results generalize previous exact results obtained for special polymer
observables \cite{tian22} and provide an analytical framework for distance
fluctuations in proteins and other network-forming soft-matter systems. As a
first step toward the distance dynamics, we derive an exact coupled GLE for two
tagged beads in the ENM, which, to our knowledge, has yet to be obtained
explicitly even for simple linear polymer models (see
Refs.~\cite{debacco14,lim25} for related two-bead descriptions).

\textit{Elastic network model.} In this Letter, we investigate a dynamical ENM
governed by an overdamped Langevin equation
\begin{equation}
  \label{e.gnm.1}
  \gamma_m \frac{d\bm{R}_m}{dt}
  = \sum_{n=1}^N k_{mn}
  \left[(\bm{R}_n-\bm{R}_n^0) - (\bm{R}_m-\bm{R}_m^0)\right]
  + \bm{\xi}_m^0(t),
\end{equation}
where $m = 1,\dots, N$. The three-dimensional vector $\bm{R}_m(t)$ represents
the position of the $m$th bead, and $\bm{R}_m^0$ is its equilibrium
position. The parameter $\gamma_m$ denotes the friction coefficient of bead $m$.
The $m$th and $n$th beads are connected by a harmonic spring with stiffness
$k_{mn}$, where $k_{mn}=k_{nm}$ and $k_{nn}=0$.  In protein applications, each
bead may represent an amino-acid residue, and residue-dependent friction
coefficients may be introduced to account for nonuniform solvent coupling
\cite{copperman15}.

The last term on the right-hand side of Eq.~(\ref{e.gnm.1}), $\bm{\xi}_m^0(t)$,
is three-dimensional Gaussian white noise that satisfies the
fluctuation-dissipation relation (FDR) \cite{kubo91}
\begin{equation}
  \label{e.<xil-xim>}
  \left\langle \bm{\xi}_m^0(t) \bm{\xi}_n^0(t') \right\rangle
  =2k_BT \gamma_m \delta_{mn}\delta (t'-t) I_3,
\end{equation}
where $k_B$ is the Boltzmann constant, $T$ is the temperature, $I_n$ is the
$n\times n$ identity matrix, $\delta_{mn}$ is the Kronecker delta, and
$\delta(t)$ is the Dirac delta function.

We define $\bm{r}_m$ as the displacement from equilibrium,
$\bm{r}_m := \bm{R}_m - \bm{R}_m^0$. Then, Eq.~(\ref{e.gnm.1}) can be rewritten
as
\begin{equation}
  \label{e.enm.1}
  \frac{d\bm{r}_m}{dt}
  =
  \frac {1}{\gamma_m}
  \sum^N_{n=1} k_{mn}(\bm{r}_n-\bm{r}_m)
  +
  \frac {1}{\gamma_m}
  \bm{\xi}_m^0(t).
\end{equation}
Let us define an interaction matrix $L_0$ by its $(m,n)$ entry, $l_{mn}^0$, as
$l_{mn}^0 := \delta_{mn} d_n - k_{mn}$ with $d_n := \sum^N_{m=1} k_{mn}$. We also
define a mobility matrix $H$ by its $(m,n)$ entry, $h_{mn}$, as
$h_{mn} = \gamma_m^{-1} \delta_{mn}$. Both $L_0$ and $H$ are symmetric
$N \times N$ matrices.
We employ the supervector notation $\bm{r} := (\bm{r}_1, \dots , \bm{r}_N )$ and
$\bm{\xi}^0 := (\bm{\xi}_1^0, \dots , \bm{\xi}_N^0 )$. Then, Eq.~(\ref{e.enm.1})
can be rewritten as
\begin{equation}
  \label{e.enm.sv}
  \frac{d\bm{r}}{dt}= - L\cdot\bm{r}
  + \bm{\xi}(t),
\end{equation}
where $L := HL_0$ and $\bm{\xi}(t):= H\cdot\bm{\xi}^0(t)$. Thus, due to the
heterogeneous friction, $L$ becomes nonsymmetric \footnote{If the friction is
  homogeneous, $\gamma_m=\gamma_0 \;(m=1,\dots,N)$, then $H=\gamma_0^{-1} I_N$
  and hence $L$ is symmetric. In this case, the subsequent analysis becomes much
  simpler because there is no need to distinguish between row and column
  vectors.}. The dot denotes the contraction of a matrix with a vector and of
one vector with another.  With this supervector notation, the FDR in
Eq.~(\ref{e.<xil-xim>}) is rewritten as
\begin{equation}
  \label{e.fdr_xixi}
  \left\langle \bm{\xi}(t)\bm{\xi}(t') \right\rangle
  =2k_BT \delta (t'-t)HI_3,
\end{equation}
where $HI_3$ denotes the tensor product of $H$ in the bead-index space and $I_3$
in the spatial-coordinate space.


Suppose that the $i$th and $j$th beads are tagged ($i<j$ is assumed). We define
a reduced vector $\bm{r}''$ by removing the $i$th and $j$th entries, $\bm{r}_i$
and $\bm{r}_j$, from $\bm{r}$. Similarly, we define the reduced matrices $L''$
and $L''_0$ of order $N-2$ by removing the $i$th and $j$th rows and columns from
$L$ and $L_0$, respectively.
We assume throughout that the elastic network is connected. Accordingly, $L_0$
has one zero eigenvalue associated with uniform translation and $N-1$ positive
eigenvalues, and $L''_0$ is positive definite.
The $k$th column and $k$th row vectors of $L$ are denoted by
$\bm{l}_{\bullet k}$ and $\bm{l}_{k \bullet}$, respectively. Then, the matrix
$L$ can be expressed as $L = [\bm{l}_{\bullet1}, \dots, \bm{l}_{\bullet
  N}]$. It follows that $L''$ can be written explicitly as
$L'' = [\bm{l}_{\bullet1}'',\dots,\bm{l}_{\bullet i-1}'', \bm{l}_{\bullet
  i+1}'',\dots,\bm{l}_{\bullet j-1}'', \bm{l}_{\bullet
  j+1}'',\dots,\bm{l}_{\bullet N}'']$.


\textit{Derivation of two-bead GLE.} We now derive the two-bead GLE for the
elastic network model defined by Eq.~(\ref{e.enm.sv}). To this end, we follow
the basic procedure introduced by Zwanzig \cite{zwanzig01book}, which partitions
the full system into the system of interest and the environment to be
eliminated, instead of employing the projection-operator formalism \cite{xing11,
  miyaguchi24}. Removing the $i$th and $j$th components from
Eq.~(\ref{e.enm.sv}) [see the Supplemental Material (SM) \cite{SM25} for a
derivation], we have
\begin{equation}
  \frac{d\bm{r}''}{dt}
  \label{e.r'_eom}
  =  -{L''}\cdot(\bm{r}''-\bm{r}_{\mathrm{G}})+\bm{\xi}''(t).
\end{equation}
This is the equation of motion for the environment to be eliminated below. Here
$\bm{r}_{\mathrm{G}}(t)$ is defined by
\begin{equation}
  \label{e.r_G}
  \bm{r}_{\mathrm{G}}(t):= -L''^{-1}\cdot(\bm{l}_{\bullet i}''\bm{r}_i +
  \bm{l}_{\bullet j}''\bm{r}_j),
\end{equation}
with $L''^{-1}$ being the inverse of $L''$. Note that the expressions such as
$\bm{l}_{\bullet i}'' \bm{r}_i$ are tensor products of $(N-2)$- and
$3$-dimensional vectors.

The equations of motion for the $i$th and $j$th beads, which constitute the
system of interest, are given by (see the SM \cite{SM25})
\begin{equation}
  \label{e.ri_eom}
  \frac{d\bm{r}_{\alpha}}{dt}
  =
  -\bm{l}''_{\alpha\bullet}\cdot(\bm{r}'' - \bm{r}_{\mathrm{G}})
  +
  \frac {\tilde{k}_{ij}}{\gamma_{\alpha}}
  \left(\bm{r}_{\bar{\alpha}} - \bm{r}_{\alpha}\right)
  + \bm{\xi}_{\alpha}(t),
\end{equation}
with $(\alpha, \bar{\alpha})=(i,j)$ or $(j,i)$. The effective stiffness
$\tilde{k}_{ij}$ between the two tagged beads is defined by
\begin{equation}
  \label{e.tilde_kij}
  \tilde{k}_{ij}
  :=
  k_{ij}
  +
  \bm{l}''_i\cdot L_0''^{-1} \cdot \bm{l}''_j,
\end{equation}
where $\bm{l}''_{\alpha}$ ($\alpha=i,j$) is the vector obtained by removing the
$i$th and $j$th entries from the $\alpha$th column or row of $L_0$ (note that
$L_0$ is symmetric). The first term on the right-hand side of
Eq.~(\ref{e.tilde_kij}), $k_{ij}$, is the original stiffness, corresponding to
the direct interaction that appears in Eq.~(\ref{e.enm.1}). By contrast, the
second term is an effective contribution arising from indirect interactions
mediated by the other beads. Note also that $\tilde{k}_{ij} = \tilde{k}_{ji}$
because $L_0''^{-1}$ is symmetric.

An alternative expression for $\tilde{k}_{ij}$ can be obtained from the
determinant of the Schur complement. In fact, we have (see the SM \cite{SM25})
\begin{equation}
  \label{e.tilde_kij_2}
  \tilde{k}_{ij}
  =
  \frac { \mathrm{det}L_0'}{\mathrm{det} L_0''},  
\end{equation}
where $L_0'$ denotes the matrix obtained by removing the $i$th row and column
from $L_0$.
The right-hand side of Eq.~(\ref{e.tilde_kij_2}) can also be rewritten as
$1/(L_0'^{-1})_{jj}$ by using the standard formula for matrix inversion.
Therefore, we have $k_BT I_3 /\tilde{k}_{ij} = k_BT (L_0'^{-1})_{jj} I_3$, which
is equal to the covariance of the distance vector $\bm{r}_j - \bm{r}_i$, because
$k_BT L_0'^{-1} I_3$ is the covariance matrix of the set of vectors
$\bm{r}_k - \bm{r}_i\,(k\neq i)$ \cite{miyaguchi24}. Thus, we confirm the
equipartition relation
$\tilde{k}_{ij}\langle (\bm{r}_j - \bm{r}_i)(\bm{r}_j - \bm{r}_i) \rangle/2 =
k_BT I_3/2$ with the effective stiffness $\tilde{k}_{ij}$ rather than the
original stiffness $k_{ij}$.

The inner product between $\bm{l}''_{\alpha\bullet}$ and a formal solution of
Eq.~(\ref{e.r'_eom}) is given by
\begin{equation}
  \label{e.r'_formal_sol}
  \bm{l}''_{\alpha\bullet}\cdot \delta\bm{r}''(t)
  =
  -\bm{\xi}_{\alpha}^{\mathrm{r}}(t)
  - \int_0^t \bm{l}''_{\alpha\bullet} \cdot e^{-L''(t-\tau)}
  \cdot \dot{\bm{r}}_{\mathrm{G}} (\tau) d\tau,
\end{equation}
where $\delta\bm{r}''(t)$ is defined as
$\delta \bm{r}''(t) := \bm{r}''(t) - \bm{r}_{\mathrm{G}}(t)$ and
$\bm{\xi}_{\alpha}^{\mathrm{r}}$ is colored noise defined as
\begin{equation}
  \label{e.xi^r}
  \bm{\xi}_{\alpha}^{\mathrm{r}}(t)
  :=
  -\bm{l}_{\alpha\bullet}''\cdot e^{-L''t} \cdot \delta\bm{r}''(0) 
  -
  \bm{l}_{\alpha\bullet}''\cdot \int_0^t 
  e^{-L''(t-\tau)}\cdot \bm{\xi}''(\tau)d\tau.
\end{equation}
Substituting Eq.~(\ref{e.r'_formal_sol}) into Eq.~(\ref{e.ri_eom}) and rewriting
$\bm{r}_{\mathrm{G}}$ with Eq.~(\ref{e.r_G}) yield the two-bead GLE for the
elastic network model:
\begin{align}
  \frac{d\bm{r}_{\alpha}}{dt}
  &+
  \int_0^t
  \left[
  \mu_{\alpha\alpha}(t-\tau) \dot{\bm{r}}_{\alpha}(\tau)
  +
  \mu_{\alpha\bar{\alpha}}(t-\tau) \dot{\bm{r}}_{\bar{\alpha}}(\tau)
  \right]d\tau
  \notag\\[0.1cm]
  \label{e.gle_i}
  &=
  \frac {\tilde{k}_{ij}}{\gamma_{\alpha}}
    (\bm{r}_{\bar{\alpha}} -\bm{r}_{\alpha})
    +\bm{\xi}_{\alpha}^{\mathrm{r}} + \bm{\xi}_{\alpha},
\end{align}
where $\mu_{\alpha\beta}(t)$, with $(\alpha, \beta) = (i,i), (i, j), (j, i)$ or
$(j,j)$, is a resistance kernel defined by
\begin{equation}
  \label{e.mu_alphabeta}
  \mu_{\alpha\beta}(t)
  :=
  \bm{l}_{\alpha\bullet}''\cdot 
  e^{-L''t}L''^{-1}
  \cdot \bm{l}_{\bullet\beta}''.
\end{equation}
Note that the kernels $\mu_{\alpha\beta}(t)$ are independent of the bead
positions $\bm{r}_{\alpha}$, and thus Eq.~(\ref{e.gle_i}) is a coupled GLE with
coordinate-independent kernels. Moreover, the two-bead GLE in
Eq.~(\ref{e.gle_i}) satisfies the FDR (see the SM \cite{SM25})
\begin{equation}
  \label{e.FDR_two_bead}
  \left\langle
  \bm{\xi}_{\alpha}^{\mathrm{r}}(t)
  \bm{\xi}_{\beta}^{\mathrm{r}}(t') \right\rangle
  =
  \frac {k_BT}{\gamma_{\beta}}I_3 \mu_{\alpha\beta}(t'-t).
\end{equation}
A two-bead model in a similar form has been studied recently in
Ref.~\cite{lim25} in the context of diffusion in viscoelastic media.

Equations (\ref{e.tilde_kij}) and (\ref{e.xi^r})--(\ref{e.FDR_two_bead}) are the
main results of the first part of this Letter.
As shown by Eq.~(\ref{e.gle_i}), the forces exerted on the two tagged beads by
the other beads can be decomposed into an indirect potential force, a
memory-dependent resistance force, and a colored noise term.  From
Eq.~(\ref{e.mu_alphabeta}), it can be shown that $\gamma_i\mu_{ij}(0)$ is
precisely the indirect stiffness $\bm{l}''_i\cdot L_0''^{-1} \cdot \bm{l}''_j$
in Eq.~(\ref{e.tilde_kij}) (see the SM \cite{SM25}). This relation can be
understood as a consequence of linear response relations \cite{debacco14}.
Furthermore, the mutual kernels satisfy
$\gamma_i\mu_{ij}(t) = \gamma_j\mu_{ji}(t)$, which is a manifestation of the
Onsager reciprocity \cite{debacco14}. In the SM \cite{SM25}, which includes
Ref.~\cite{feller68}, we explicitly derive the kernels for two symmetric beads
($i$ and $j=N-i+1$) in the Rouse model \cite{rouse53, doi86}, and for two
arbitrary beads in a ring polymer.
%

If the two tagged beads are statistically equivalent, with $\gamma_i= \gamma_j$
and $\mu_{ii} = \mu_{jj}$, a GLE for the inter-bead distance vector
$\tilde{\bm{r}}_i := \bm{r}_i- \bm{r}_j$ can be derived from the two-bead
GLE. Specifically, setting $\mu_{\mathrm{s}} := \mu_{ii} = \mu_{jj}$,
$\mu_{\mathrm{m}}:=\mu_{ij} = \mu_{ji}$ and
$\tilde{\mu}_{\mathrm{eb}}(t):= \mu_{\mathrm{s}}(t) - \mu_{\mathrm{m}}(t)$, we
obtain a GLE for $\tilde{\bm{r}}_i$:
  \begin{equation}
  \label{e.gle_r_1N}
  \frac {d\tilde{\bm{r}}_i}{dt}
  +
  \int_0^t
  \tilde{\mu}_{\mathrm{eb}}(t-\tau)\dot{\tilde{\bm{r}}}_i(\tau) d\tau
  =
  -\frac {\tilde{k}_{ij}}{\tilde{\gamma}} \tilde{\bm{r}}_i
  +\tilde{\bm{\xi}}_{i, \mathrm{eb}}^{\mathrm{r}} + \tilde{\bm{\xi}}_i,
\end{equation}
where we set
$\tilde{\bm{\xi}}_{i, \mathrm{eb}}^{\mathrm{r}}:= \bm{\xi}_i^{\mathrm{r}} -
\bm{\xi}_j^{\mathrm{r}}$ and $\tilde{\bm{\xi}}_i:= \bm{\xi}_i - \bm{\xi}_j$.
Here, $\tilde{\gamma}^{-1} := \gamma_i^{-1} + \gamma_j^{-1}$ is an effective
friction constant, and ``$\mathrm{eb}$'' stands for ``equivalent beads''. It is
straightforward to verify that the FDRs hold for Eq.~(\ref{e.gle_r_1N}):
$\langle \tilde{\bm{\xi}}_{i, \mathrm{eb}}^{\mathrm{r}}(t)\tilde{\bm{\xi}}_{i,
  \mathrm{eb}}^{\mathrm{r}}(t') \rangle = k_BT
\tilde{\mu}_{\mathrm{eb}}(t'-t)I_3/\tilde{\gamma}$, and
$\langle \tilde{\bm{\xi}}_i(t)\tilde{\bm{\xi}}_i(t') \rangle = 2k_BT
\delta(t'-t)I_3/\tilde{\gamma}$. However, the GLE in Eq.~(\ref{e.gle_r_1N}) is
derived here under the statistical-equivalence condition specified above,
although it is useful for explicit calculations of the kernels for exactly
solvable models (see the SM \cite{SM25}). The above procedure generally fails
because the conjugate vector $\tilde{\bm{r}}_j$ [see Eq.~(\ref{e.tilde_r_i})] is
not properly projected out of the equation for $\tilde{\bm{r}}_i$. We therefore
derive below an alternative inter-bead GLE that remains valid for arbitrary bead
pairs, including those with $\mu_{ii} \neq \mu_{jj}$.

\textit{Derivation of GLE for inter-bead distance.} The distance $\ell(t)$
between two tagged beads is important for understanding the conformational
dynamics of proteins and biopolymers \cite{ayaz21, allemand06} as well as for
interpreting data obtained from experiments such as the photoinduced electron
transfer and F\"{o}rster resonance energy transfer \cite{kou04, min05,
  kuzmenkina05}. To derive the GLE for $\ell(t)$, we first investigate a
distance vector $\tilde{\bm{r}}_i$ and its conjugate $\tilde{\bm{r}}_j$ defined
by
\begin{equation}
  \label{e.tilde_r_i}
  \tilde{\bm{r}}_i := \bm{r}_i- \bm{r}_j,\quad
  \tilde{\bm{r}}_j := p_i\bm{r}_i+ p_j\bm{r}_j, \quad
  \tilde{\bm{r}}_k := \bm{r}_k\,(k\neq i, j),
\end{equation}
where $p_i$ and $p_j$ are constants to be specified below.  Using these vectors,
we define an $N$-dimensional supervector
$\tilde{\bm{r}} := (\tilde{\bm{r}}_1,\dots, \tilde{\bm{r}}_N)$. The
transformation from $\bm{r}$ to $\tilde{\bm{r}}$ can be expressed by an
$N\times N$ matrix $P$ as $\tilde{\bm{r}} = P \bm{r}$; similarly,
$\tilde{\bm{\xi}}$ is defined as $\tilde{\bm{\xi}} := P \bm{\xi}$. Moreover, an
$N\times N$ matrix $\tilde{L}$ is defined by $\tilde{L} := PL P^{-1}$ (see the
SM for explicit expressions of $P$ and $\tilde{L}$ \cite{SM25}).


We derive equations of motion for $\tilde{\bm{r}}_i$ and $\tilde{\bm{r}}_j$ from
Eq.~(\ref{e.ri_eom}). To do so, we define $\tilde{L}'$ as the matrix obtained by
eliminating the $i$th row and column from $\tilde{L}$. Similarly,
$\tilde{\bm{r}}'$ is obtained by eliminating the $i$th entry from
$\tilde{\bm{r}}$. Setting $p_i = \gamma_i/(\gamma_i+\gamma_j)$ and
$p_j = \gamma_j/(\gamma_i+\gamma_j)$, and using Eq.~(\ref{e.ri_eom}) with
$\alpha = i$ and $\alpha=j$, we obtain
\begin{equation}
  \label{e.eom.tilde_r_j}
  \frac {d{\tilde{\bm{r}}}_j}{dt}
  =
  - \tilde{\bm{l}}_{j\bullet}'' \cdot(\tilde{\bm{r}}'' - \bm{r}_{\mathrm{G}})
  + \tilde{\bm{\xi}}_j
  =
  - \tilde{\bm{l}}_{j\bullet}' \cdot (\tilde{\bm{r}}' - \tilde{\bm{r}}_{\mathrm{G}})
  + \tilde{\bm{\xi}}_j,
\end{equation}
where $\tilde{\bm{l}}_{j\bullet}$ is the $j$th row of $\tilde{L}$, and
$\tilde{\bm{r}}_{\mathrm{G}}$ is defined by inserting $\tilde{\bm{r}}_j$ into
$\bm{r}_{\mathrm{G}}$ [Eq.~(\ref{e.r_G})] as the $j$th entry; that is,
$(\tilde{\bm{r}}_{\mathrm{G}})_j = \tilde{\bm{r}}_j$, while the other elements
are the same as those of $\bm{r}_{\mathrm{G}}$. From Eqs.~(\ref{e.r'_eom}) and
(\ref{e.eom.tilde_r_j}), we have
\begin{equation}
  \frac{d\tilde{\bm{r}}'}{dt}
  \label{e.tilde_r'_eom}
  =
  -{\tilde{L}'}\cdot(\tilde{\bm{r}}'-\tilde{\bm{r}}_{\mathrm{G}})
  +\tilde{\bm{\xi}}'(t),
\end{equation}
where $\tilde{\bm{r}}'$ and $\tilde{\bm{\xi}}'$ are obtained from
$\tilde{\bm{r}}$ and $\tilde{\bm{\xi}}$ by removing the $i$th entry. Thus,
Eq.~(\ref{e.tilde_r'_eom}) describes the environment to be eliminated. The
equation for the variable of interest, $\tilde{\bm{r}}_i$, is obtained by
subtracting Eq.~(\ref{e.ri_eom}) with $\alpha = j$ from the same equation with
$\alpha = i$:
\begin{equation}
  \label{e.eom.tilde_r_i}
  \frac {d{\tilde{\bm{r}}}_i}{dt}
  =
  - \tilde{\bm{l}}_{i\bullet}' \cdot (\tilde{\bm{r}}' - \tilde{\bm{r}}_{\mathrm{G}})
  - \frac {\tilde{k}_{ij}}{\tilde{\gamma}} \tilde{\bm{r}}_i + \tilde{\bm{\xi}}_i,
\end{equation}
where $\tilde{\gamma}^{-1} = \gamma_i^{-1} + \gamma_j^{-1}$ as before.

\begin{figure}[t]
  \begin{minipage}[bt]{5.2cm}
    \includegraphics[width=5.2cm]{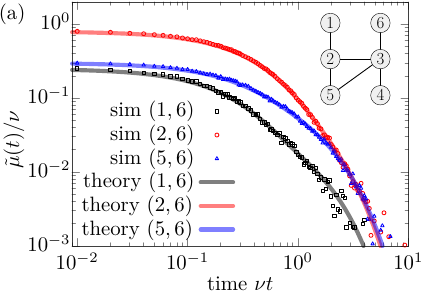}
  \end{minipage}
  \begin{minipage}[bt]{3.3cm}
    \includegraphics[width=3.3cm]{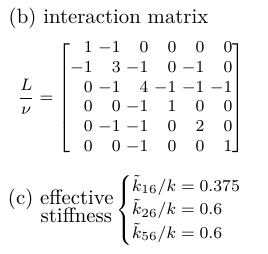}
    \vspace*{.2cm}
  \end{minipage}
  \caption{\label{f.toy6} (a) Memory kernels $\tilde{\mu}(t)$ for the bead pairs
    $(i,j) = (1,6)$, $(2,6)$, and $(5,6)$ in the ENM shown in the inset. The
    solid lines are calculated from Eq.~(\ref{e.tilde_mu}), while the symbols
    show numerical estimates of the memory kernels obtained from trajectory data
    generated by Eq.~(\ref{e.enm.sv}) [see the SM \cite{SM25}].  (b) The
    corresponding interaction matrix $L$, where $\nu = k/\gamma$.  (c)
    The effective stiffnesses for these bead pairs.}
  
\end{figure}
More precisely, $\tilde{\bm{r}}_{\mathrm{G}}$ is given by
$\tilde{\bm{r}}_{\mathrm{G}} = \bm{1}' \tilde{\bm{r}}_j - \tilde{L}'^-\cdot
\tilde{\bm{l}}_{\bullet i}' \tilde{\bm{r}}_i$, where $\tilde{L}'^-$ is a
generalized inverse of $\tilde{L}'$ \cite{horn12book} (note that $\tilde{L}'$ is
singular; see the SM \cite{SM25}), and the $N$-dimensional vector $\bm{1}$ is
defined by $\bm{1}:=(1,\dots,1)$ (i.e., all its elements are unity). The matrix
$\tilde{L}'^-$ can be constructed by inserting zeros into the $j$th row and
column of $L''^{-1}$. Note also that the expressions $\bm{1}' \tilde{\bm{r}}_j$
and $\tilde{L}'^-\cdot \tilde{\bm{l}}_{\bullet i}' \tilde{\bm{r}}_i$ are tensor
products of $(N-1)$- and $3$-dimensional vectors.  It can also be shown that
$\tilde{L}'\cdot \bm{1}' = \bm{0}$ and
$\tilde{\bm{l}}'_{i\bullet}\cdot \bm{1}' = 0$ (see the SM \cite{SM25}), and
therefore, in Eqs.~(\ref{e.tilde_r'_eom}) and (\ref{e.eom.tilde_r_i}),
$\tilde{\bm{r}}_{\mathrm{G}}$ can be replaced with
\begin{equation}
  \label{e.tilde_rG}
  \tilde{\bm{r}}_{\mathrm{G}}
  =
  - \tilde{L}'^- \cdot \tilde{\bm{l}}_{\bullet i}'\tilde{\bm{r}}_i.
\end{equation}

By solving Eq.~(\ref{e.tilde_r'_eom}) and substituting the result, together with
Eq.~(\ref{e.tilde_rG}), into Eq.~(\ref{e.eom.tilde_r_i}), we obtain the GLE for
the inter-bead distance vector $\tilde{\bm{r}}_i$:
\begin{equation}
  \label{e.gle-inter-bead}
  \frac {d{\tilde{\bm{r}}}_i}{dt}
  +
  \int_0^t \tilde{\mu} (t-\tau) \dot{\tilde{\bm{r}}}_i (\tau)d\tau
  =
  - \frac {\tilde{k}_{ij}}{\tilde{\gamma}} \tilde{\bm{r}}_i
  + \tilde{\bm{\xi}}_i^{\mathrm{r}} +\tilde{\bm{\xi}}_i,
\end{equation}
This equation has exactly the same form as Eq.~(\ref{e.gle_r_1N}); however,
Eq.~(\ref{e.gle-inter-bead}) is valid for an arbitrary bead pair $(i, j)$, and
the definitions of the memory kernel $\tilde{\mu}(t)$ and the colored noise
$\tilde{\bm{\xi}}^{\mathrm{r}}_i(t)$ differ completely from those in
Eq.~(\ref{e.gle_r_1N}). Specifically, they are given by
\begin{align}
  \label{e.tilde_mu}
  \tilde{\mu}(t)
  &:=
  \tilde{\bm{l}}_{i\bullet}' \cdot e^{- \tilde{L}'t}\tilde{L}'^-
  \cdot \tilde{\bm{l}}_{\bullet i}',
  \\[0.1cm]
  \label{e.tilde_xi_i^r}
  \tilde{\bm{\xi}}_i^{\mathrm{r}}
  &:=
  -\tilde{\bm{l}}_{i\bullet}' \cdot e^{-\tilde{L}'t} \cdot \delta\tilde{\bm{r}}'(0) 
  -\tilde{\bm{l}}_{i\bullet}'\cdot
  \int_0^t e^{-\tilde{L}'(t-\tau)}\cdot \tilde{\bm{\xi}}'(\tau)d\tau,
\end{align}
where $\delta\tilde{\bm{r}}':= \tilde{\bm{r}}' - \tilde{\bm{r}}_{\mathrm{G}}$.
Figure~\ref{f.toy6} shows the memory kernels $\tilde{\mu}(t)$ for a six-bead
ENM, calculated from Eq.~(\ref{e.tilde_mu}). The results show that the kernel
depends on the choice of the tagged pair.
The fluctuation-dissipation relation also holds (see the SM \cite{SM25}):
\begin{equation}
  \label{e.FDR.inter-bead-vector}
  \bigl\langle
  \tilde{\bm{\xi}}_i^{\mathrm{r}}(t)
  \tilde{\bm{\xi}}_i^{\mathrm{r}}(t')
  \bigr\rangle
  =
  \frac {k_BT}{\tilde{\gamma}} I_3{\tilde{\mu}}(t'-t).
\end{equation}

Next, we derive a GLE for the inter-bead scalar distance
$\ell = |\bm{R}_i-\bm{R}_j|$ in the ENM defined by Eq.~(\ref{e.gnm.1}). We
denote the inter-bead distance vector by $\bm{\ell} := \bm{R}_i -\bm{R}_j$, and
its equilibrium counterpart by $\bm{\ell}_0 := \bm{R}_i^0 -\bm{R}_j^0$. We
assume that the displacement $|\bm{\ell} - \bm{\ell}_0| = |\tilde{\bm{r}}_i|$ is
small compared with the equilibrium distance
$\ell_0 = |\bm{R}_i^0- \bm{R}_j^0|$. Then, $\ell$ is approximated as $ \ell^2
\approx \ell_0^2 ( 1 + 2 \hat{\bm{\ell}}_0\cdot \tilde{\bm{r}}_i / \ell_0)$,
where $\hat{\bm{\ell}}_0 := \bm{\ell}_0 / |\bm{\ell}_0|$. Therefore, we have
\begin{equation}
  \label{e.ell.approx}
  \ell \approx \ell_0 + \hat{\bm{\ell}}_0\cdot \tilde{\bm{r}}_i.
\end{equation}
Moreover, differentiating Eq.~(\ref{e.ell.approx}) with respect to $t$, we
obtain $\dot{\ell} \approx \hat{\bm{\ell}}_0\cdot \dot{\tilde{\bm{r}}}_i$.

Taking the contraction of Eq.~(\ref{e.gle-inter-bead}) with $\hat{\bm{\ell}}_0$,
we obtain the GLE for the inter-bead distance:
\begin{equation}
  \label{e.gle.ell}
  \frac {d\ell}{dt}
  + \int_0^t \tilde{\mu} (t-\tau) \dot{\ell}(\tau) d\tau
  =
  - \frac {\tilde{k}_{ij}}{\tilde{\gamma}} (\ell-\ell_0)
  + \xi_{\ell}^{\mathrm{r}} + \xi_{\ell},
\end{equation}
where
$\xi_{\ell}^{\mathrm{r}} := \hat{\bm{\ell}}_0\cdot
\tilde{\bm{\xi}}_i^{\mathrm{r}}$, and
$\xi_{\ell} := \hat{\bm{\ell}}_0\cdot \tilde{\bm{\xi}}_i$.
The FDR for Eq.~(\ref{e.gle.ell}),
$\langle \xi_{\ell}^{\mathrm{r}}(t)\xi_{\ell}^{\mathrm{r}}(t') \rangle = k_BT
\tilde{\mu}(t'-t)/\tilde{\gamma}$, follows directly from
Eq.~(\ref{e.FDR.inter-bead-vector}). Remarkably, the memory kernel $\tilde{\mu}$
for the inter-bead distance is identical to that for the inter-bead distance
vector [Eq.~(\ref{e.tilde_mu})]. Equations
(\ref{e.gle-inter-bead})--(\ref{e.FDR.inter-bead-vector}) and (\ref{e.gle.ell})
constitute the main results of the second part of this Letter.


\textit{Conclusion.} In this Letter, we derived exact GLEs for pair coordinates
in overdamped elastic networks. Specifically, we obtained an exact coupled GLE
for two tagged beads, an exact GLE for their relative coordinate, and, within
the small-displacement approximation, a GLE for the inter-bead distance. The
corresponding memory kernels are determined by the reduced interaction matrices
$L''$ and $\tilde{L}'$, providing an explicit projection from high-dimensional
network dynamics onto a small set of pair observables.

These results extend GLE descriptions beyond single-bead motion and generalize
previous exact results that were limited to special polymer observables. They
therefore provide a general analytical framework for constructing
low-dimensional dynamics with memory in harmonic network systems.

These theoretical results are illustrated for a simple six-bead network
[Fig.~\ref{f.toy6}], and for the Rouse model and the ring polymer [Figs.~S1 and
S2 in the SM]. The memory kernels for the Rouse model and ring polymer exhibit a
power-law-like decay over an intermediate-time regime, whereas no such slow
decay is observed for the six-bead network. This difference may be due mainly to
the much smaller system size of the latter ($N=6$, compared with $N=200$ for the
polymer models), which greatly reduces the number of relaxation modes and the
range of relaxation time scales.  Network connectivity may also affect the
relaxation spectrum; for example, as shown in the SM, adding a positive spring
bond to a Rouse chain strictly shortens its longest relaxation time.

The present theory also suggests a practical route toward applications. Once the
interaction matrix $L_0$ and a friction model $H$ are specified, the GLE for an
arbitrary tagged pair in a dynamical ENM can be constructed explicitly. For
example, $L_0$ for proteins can be built from X-ray and NMR structural data
using a distance-based cutoff rule \cite{tirion96,bahar98}, whereas $L_0$ for
chromatin can be inferred from high-throughput chromosome conformation capture
(Hi-C) data \cite{shinkai20}. To describe slow distance fluctuations in proteins
observed in experiments such as photoinduced electron transfer
\cite{kou04,min05} and F\"orster resonance energy transfer \cite{kuzmenkina05},
the present harmonic description may need to be extended to incorporate
additional slow physics, such as rugged energy landscapes
\cite{yang03,xing06,goychuk18} or fluctuating diffusivity \cite{miyaguchi22,
  akimoto26rev}.  Such extensions may provide efficient coarse-grained
descriptions of distance fluctuations in proteins and other network-forming
soft-matter systems. It is also interesting to utilize the present framework to
study mechanical responses of disordered materials under external stress
\cite{zaccone23book}.





\textit{Acknowledgments.}  S.S. was supported by JSPS KAKENHI Grant No.
JP23H04297 and JST CREST Grant No. JPMJCR23N3. T.M. was supported by JSPS
KAKENHI Grant Nos. JP22K03436 and JP26K06927.

\textit{Data availability.}  The data that support the findings of this article
were generated by numerical simulations. The source code and parameters used to
generate the simulations, together with the code used for the theoretical
calculations, are publicly available in Ref.~\cite{github-tm26}.



%

\end {document}